\documentclass{article}
\usepackage{spconf,amsmath,graphicx}
\usepackage{xcolor}
\usepackage{cite}
\usepackage{url}
\usepackage{color}
\usepackage{booktabs}
\usepackage{multirow}
\usepackage{subfigure}
\usepackage{tabularx}

\let\OLDthebibliography\thebibliography
\renewcommand\thebibliography[1]{
  \OLDthebibliography{#1}
  \setlength{\parskip}{0pt}
  \setlength{\itemsep}{0pt plus 0.5ex}
}

\title{Towards Automatic Transcription of Polyphonic Electric Guitar Music: \\A new Dataset and A Multi-loss Transformer Model}
\name{Yu-Hua Chen$^{1,2,3}$, Wen-Yi Hsiao$^1$, Tsu-Kuang Hsieh$^1$, Jyh-Shing Roger Jang$^3$, and Yi-Hsuan Yang$^{1,2}$}
\address{$^1$ Taiwan AI Labs, Taiwan,~
  $^2$ Academia Sinica, Taiwan,~
  $^3$ National Taiwan University, Taiwan}
\begin{document}
\ninept
\maketitle
\begin{abstract}

In this paper, we propose a new dataset named EGDB, that contains transcriptions of the electric guitar performance of 240 tablatures rendered with different tones. Moreover, we benchmark the performance of two well-known transcription models proposed originally for the piano on this  dataset, along with a multi-loss Transformer model that we newly propose. Our evaluation on this dataset and a separate set of real-world recordings demonstrate the influence of timbre on the accuracy of guitar sheet transcription, the potential of using multiple losses for Transformers, as well as the room for further improvement for this task.
\end{abstract}
\section{Introduction}
\label{sec:intro}

Automatic music transcription (AMT) is a fundamental task in audio signal processing and music information retrieval (MIR) \cite{emiya2010multipitch,benetos2013automatic,su2015escaping,su15taslp,hawthorne2018enabling,benetos2018automatic,thickstun2018invariances,omnizart}.  
Among musical instruments, the AMT of piano music has received the most attention in the literature \cite{vincent2010adaptive,lee12tmm,sigtia16taslp,cogliati2016transcribing,kelz2016potential,hawthorne2017onsets,kim2019adversarial,TTtranscription,hawthorne2021sequence}, 
partly thanks to the availability of large-scale labeled datasets such as MAPS \cite{emiya2010multipitch} and MAESTRO ($\sim$200 hours) \cite{hawthorne2018enabling}.
Creating labeled data for piano transcription has been found feasible,
because the labels can be entered while playing the music using a 
MIDI keyboard, or a piano with note-capturing devices
\cite{benetos2013automatic,su2015escaping}.
Saving piano scores as MIDI files also comes handy, for a MIDI file can  specify the onset, offset, velocity of notes, and the usage of  piano pedals.
Other instruments, 
however, do not have such an advantage, due to the differences in the physical mechanisms creating the sounds. 
Without the support of large labeled datasets, researchers cannot easily employ state-of-the-art machine learning algorithms to promote the progress of AMT for instruments other than the piano.

This paper aims at contributing to the AMT of music played by an electric guitar, which plays an important role in many musical genres such as rock and metal. Guitar transcription is technically challenging as it entails a similar polyphonic transcription problem as piano transcription. 
Our contributions are threefold.

First, extending the methodology proposed by Xi \emph{et al.} \cite{xi2018guitarset}, we propose a new dataset for electric guitar transcription.
Xi \emph{et al.} proposed to attach a special hexaphonic pickup \cite{ogrady09iet,mignecoPHD12} to each string of an acoustic guitar to capture activity signals per string. The hexaphonic recordings were then analyzed using a semi-automatic approach combining the pYIN monophonic pitch estimator \cite{pyin} and manual inspections to generate annotations for each string individually. They used a microphone to record the sounds of the acoustic guitar as the audio counterpart of the annotations. The resulting dataset, named GuitarSet \cite{xi2018guitarset}, is rather small and comprises the annotations of 60 unique pieces
(totaling 30 minutes) from a single guitar.  
In contrast, we collect the hexaphonic recordings from an electric guitar, and use a JUCE program \cite{juce}
to control a digital audio workstation (DAW) to automatically re-render the audio recordings of the ``Direct Input'' (DI) using different amplifiers (Amps), including low-gain Amps and high-gain ones. 
Moreover, we employ a new collecting pipeline so as to reduce the effort of manual inspection. 
The final dataset contains six copies of 118 minutes of guitar playing, each copy being associated with a different timbre. 

The new dataset, named ``EGDB,'' 
is constructed in this way to account for the diverse timbre associated with electric guitar.
Having multiple guitar tones 
makes it possible to test a trained model on held-out unseen tones for generalizability. 
We make the dataset publicly available. Link can be found at our demo website:  \url{https://ss12f32v.github.io/Guitar-Transcription/}.



Our second contribution  lies in the development of a new AMT model using Transformers \cite{vaswani2017attention}.
Very recently, Hawthorne \emph{et al.} \cite{hawthorne2021sequence} proposed a sequence-to-sequence (seq2seq) piano transcription model with Transformers, attaining the state-of-the-art accuracy on  MAESTRO \cite{hawthorne2018enabling}.
While previous work on AMT, such as the iconic ``onset and frames'' (OAF) model \cite{hawthorne2017onsets}, tends to optimizes their network using frame-wise losses, this particular model \cite{hawthorne2021sequence} measures loss only in the symbolic space at the output of the Transformer decoder. In light of the possible complementary advantages of both loss functions, we propose an extension of this seq2seq model \cite{hawthorne2021sequence} by employing both losses from encoder and decoder.
We will open source the code of this new model in the near future.


As the final contribution, we benchmark the performance of our implementation of the OAF model \cite{hawthorne2017onsets}, the seq2seq Transformer model \cite{hawthorne2021sequence}, and the proposed multi-loss seq2seq Transformer model for guitar sheet transcription over EGDB, considering both the scenarios where the timbre of the test data is seen or unseen during training.
Moreover, we report an additional experiment transcribing real-world guitar recordings downloaded from YouTube.
Our experiments validate the importance of considering multiple guitar timbres, and the effectiveness of adopting multiple losses. 

\section{Related Work}
\label{sec:related}

Guitar transcription differs from piano transcription in many aspects, including the diversity of guitar tones, the intensive use of guitar playing techniques \cite{Abesser2010icassp,Reboursiere2012,kehling2014automatic,su2014sparse,chen15ismir,su2019tent} such as bending and sliding, as well as the need of string (fingering) detection (i.e., to detect which string a note is played) \cite{barbancho10taslp,wiggins2019guitar}.
We differentiate below two types of guitar transcription tasks: \emph{tab transcription}, which involves detecting not only the pitches but also the string playing the pitches, and \emph{sheet transcription}, which does not deal with string detection. 
We focus on the latter in this paper, leaving the former as a future work.

Existing datasets for guitar transcription are rather small in size.
For example, the IDMT-SMT-GUITAR database \cite{kehling2014automatic} contains only 17 unique
guitar licks with groundtruth labels for building transcription models.
The GuitarSet \cite{xi2018guitarset} provides only 30 unique comping tracks and 30 solo tracks, in only one timbre (despite that the performance of multiple guitarists over these 60 tracks are recorded).
The guitar playing techniques database \cite{su2014sparse} is solely for the detection of playing techniques, not the estimation of pitches and onset times.
As a result, progress in guitar transcription has been slower than that of piano transcription, with few deep learning based attempts to date.

To our knowledge, the TabCNN model developed by Wiggins and Kim \cite{wiggins2019guitar} may represent the state-of-the-art for guitar tab transcription. The model, which is fully-convolutional, is trained on the GuitarSet to estimate guitar tablature from audio of a solo acoustic guitar. 
We do not benchmark the performance of this model using our dataset, since our focus here is on sheet transcription. 
Moreover, in the context of piano transcription, it has been shown  that models with recurrent layers tend to outperform fully-convolutional models \cite{sigtia16taslp}. 
While this needs to be further tested in the future, we consider in this work only recurrent and Transformer-based methods, which have been widely used in piano transcription.




Given an audio recording, the OAF model \cite{hawthorne2017onsets} uses separate but jointly-trained
convolutional recurrent neural network heads for detecting note onsets, note velocities, and frame-wise note presence, respectively. 
The training objective is to minimize the frame-wise cross-entropy loss for onsets and velocity at specific frames (where there is an onset, according to groundtruth labels) and the frame-wise cross-entropy loss for note presence across all the frames.
Kim and Bello \cite{kim2019adversarial} improved the performance of OAF by adding a fully convolutional discriminator discriminating real and generated piano rolls in an adversarial setting \cite{gan}. 

Recently, the use of Transformers \cite{vaswani2017attention} for AMT has been shown possible for piano transcription by Hawthorne \emph{et al.} \cite{hawthorne2021sequence}, who used a generic encoder-decoder Transformer architecture and considered the output of the decoder, i.e., a sequence of note-level ``tokens'', such as \texttt{NOTE-ON} and \texttt{NOTE-OFF} 
\cite{oore2018time} (see Section \ref{sec:model:LM} for more details), as the result of transcription. 
Specifically, the encoder of the model takes the spectrogram of an audio segment as input and converts each frame into a hidden representation via a number of bidirectional self-attention layers. 
The groundtruth transcription result of the audio segment is represented as a sequence of tokens beforehand while preparing the training data, and is treated as the target output of the decoder of the model.
The decoder learns to autoregressively generate the target sequence by self-attending to tokens previously-generated by itself as well as cross-attending to the hidden states of all the input audio frames.
The encoder and decoder are jointly optimized by minimizing the reconstruction error of the target sequence, measured by the cross-entropy (CE) loss at each timestep of the sequence using the teach-forcing training strategy. In this paper, we also refer to this model as an ``CE-only Transformer,'' for its training process considers only the CE loss in the space of symbolic tokens at the decoder side.


Transformer decoders have been employed in the context of automatic guitar tab generation \cite{chen2020automatic,sarmento2021dadagp}, which learns to generate original music from a large collection of guitar tabs (without the corresponding audio files). It might be possible to use such tabs to pre-train the decoder of our model, but we leave this as a future work.


\section{The Proposed Dataset}
\label{sec:dataset}


\subsection{Audio Collection}
\label{sec:dataset:audio}
Following Xi \emph{et al.} \cite{xi2018guitarset}, we employ a hexaphonic pickup to capture audio from each string individually. The pickup is attached to a Stratocaster-type electric guitar by an experienced technician. 
Unlike the case of using a microphone to collect the sound of an \emph{acoustic guitar} as done by Xi \emph{et al.} \cite{xi2018guitarset}, we opt for collecting a clean ``Direct Input'' (DI) sound from our \emph{electric guitar} using a DI box \cite{di}.
Specifically, we use high-quality Mogami cables (to reduce possible noises cause by equipment) to connect the hexaphonic pickup to a 
DI box with six output jacks, and then to the audio interface on a computer. The summation of the signals from the six output jacks (one corresponding to each string) can then be treated as an audio waveform, i.e., the \emph{DI recording}.
Using the DI reduces the effect of the recording environment and makes it easier to further apply Amps to re-render the sounds.

We collect over 500 guitar tabs encompassing compositions of solo, arpeggios and comping in various genres. 
The third author of the paper, who is a professional guitarist playing in a post-rock band, spends almost two months using that electric guitar to faithfully perform the tabs in a studio, following along through a headset a click track whose beat-per-minute (BPM) is set according to the corresponding tab. 
The throughput is roughly 5 minutes of recorded performance per working day. This yields the DI recordings of 240 unique tabs, amounting to 118 minutes worth of audio data.

\subsection{Annotation}
\label{sec:dataset:label}
Since our musician is asked to faithfully follow the reference tabs and click tracks while playing, we use the following simple automatic process to get the annotations of onsets. 
First, we estimate the \emph{expected} onset times of the notes from the tabs, using the BPM specified on the tabs to go from symbolic timing (in subbeats) to absolute timing (in milliseconds).
Second, we run an onset detection algorithm \cite{bock2012evaluating} over the recorded DI signal per string to estimate the \emph{actual} onset times of the notes. 
The expected onset times (from the tabs) and actual onset times (in the audio performances) will not be exactly the same, but for most cases there is a nice one-to-one mapping between them. 
Therefore, we can simply label the pitch of each actual onset according to that of the closest expected onset, without using sophisticated alignment algorithms.

However, the above process cannot be used for the annotation of offset times, due to the absence of a reliable offset detection algorithm. For now, we label the offset time for each note according to the actual onset time of the note ``plus'' the \emph{expected} duration (in milliseconds) of the note.


After the aforementioned annotation process, the first author of the paper, who is an amateur guitarist, manual checks the annotations for all the 240 recordings. Only very few onset labels have to be (and are) manually corrected. In contrast, the offset labels are less reliable. We leave the improvement in offset labels as a future work, using the present offset labels in model training but not evaluating the performance of the benchmarked models in offset detection.

While performing the guitar tabs, our musician employs \emph{playing techniques} such as bending when they are specified on the tabs.
It is possible to get annotations of the usage of these playing techniques by referring to the tabs, but we leave this as a future work.
Similarly, although we have access to \emph{fingering} information from individual DI signals, for simplicity, we view string detection a future work.
Moreover, we do not transcribe the note velocity.

\subsection{Timbre Re-rendering}
\label{sec:dataset:timbre}
For timbre augmentation, we use the following five commonly-used and diverse Amplifiers selected by our musician to distort and re-render the DI recordings via Guitar Rig 5 \cite{guitarrig} digitally: i) \texttt{Mesa Boogie Mark v}, ii) \texttt{Fender Twin amplifier}, iii) \texttt{Marshall JCM2000}, iv) \texttt{Roland JC120}, and v) \texttt{Marshall Plexi}.
We automate the rendering process using an in-house program written 
in JUCE, an open-source cross-platform C++ application framework \cite{juce}. 

\begin{figure}
    \centering
    \includegraphics[width=1\columnwidth]{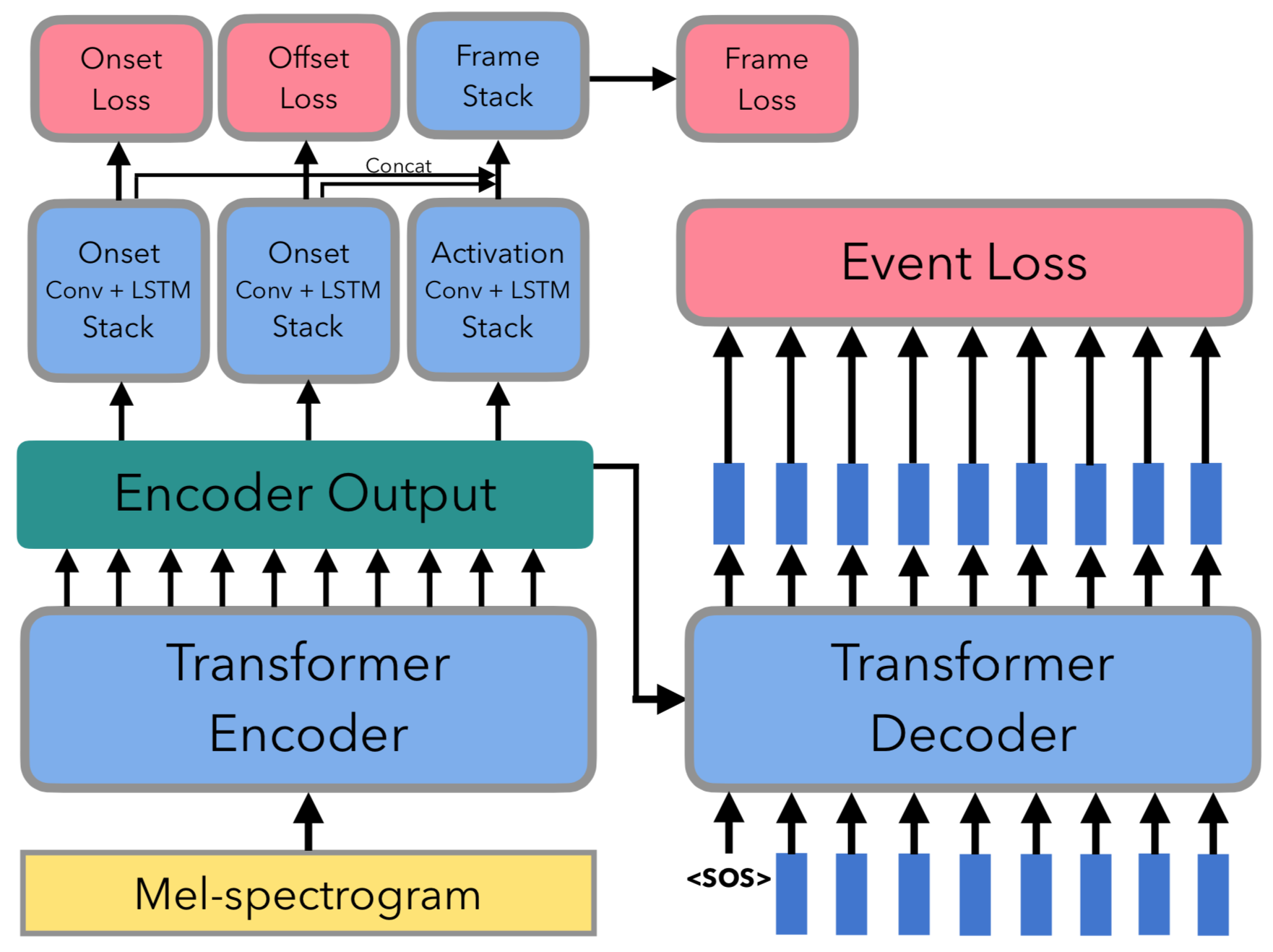}
    \\
    \caption{Schematic diagram of the proposed multi-loss seq2seq Transformer model for AMT, with additional losses at the output of the encoder compared to the model proposed by Hawthorne \emph{et al.} \cite{hawthorne2021sequence}. The input to the ``frame stack'' is a concatenation of the output from the ``onset,'' ``offset,'' and ``activation'' stacks.} 
    \label{fig:system}
\end{figure}

\section{The Proposed Transcription Model}
\label{sec:model}


Figure \ref{fig:system} shows a  diagram of the proposed multi-loss seq2seq Transformer model, which falls back conceptually to the CE-only Transformer model \cite{hawthorne2021sequence} if dropping the three loss terms associated with the encoder output. We provide the details below.


We note that, while we develop the model in the context of guitar transcription, the architecture is general and can be adapted to other AMT tasks as well (the same holds for OAF \cite{hawthorne2017onsets} and the CE-only Transformer \cite{hawthorne2021sequence}). Moreover, we can further add  string- and playing technique-related prediction heads and losses in the future.

\subsection{Transformer Encoder} 
\label{sec:model:AM}

The encoder contains primarily a stack of self-attention layers. 
Unlike \cite{hawthorne2021sequence}, we add an ``onset stack,'' an ``offset stack,'' and an ``activation stack,'' each comprising a feed-forward dense layer, to estimate the presence of note onset, presence of note offset, and the holding of a pitch for each frame using the hidden representation produced by the last self-attention layer of the encoder, as shown in Figure \ref{fig:system}. 
Moreover, we consider the following 
losses to provide supervisory signals to train the encoder.
$L_\text{frame}$ is from the ``frame stack,'' which takes the concatenation of the output of the aforementioned three stacks as input, comparing the activation probability of the 45 possible pitches for guitar (\texttt{E2} to \texttt{C6}) of the output of this stack with that of the groundtruth (which is binary and possibly multi-hot) using cross entropy.
$L_\text{onset}$ and $L_\text{offset}$ compare the probability of the presence of onset and offset of the 45 pitches with the groundtruth (also using cross entropy) at the output of the onset and offset stacks, respectively, concerning only the frames at the groundtruth onset and offset times (as done in OAF \cite{hawthorne2017onsets}).
We call the summation of the three losses as $L_\text{AM}$, namely $L_\text{AM}=L_\text{frame}+L_\text{onset}+L_\text{offset}$.

\subsection{Transformer Decoder} 
\label{sec:model:LM}

Our decoder follows pretty much the setting in \cite{hawthorne2021sequence}, representing music as ``event tokens'' using a vocabulary with 3 types of tokens: 
\begin{description}
   \item[Note] [90 possible values] Each of the 45 possible pitches from 
   \texttt{E2} to \texttt{C6}
   has its \texttt{NOTE-ON} and \texttt{NOTE-OFF} events. 
   
   \item[Time] [512 possible values] Every \texttt{NOTE-ON} or \texttt{NOTE-OFF} will be followed by a \texttt{WHICH-FRAME} event that indicates the exact frame among the input frames the note-on or note-off event takes place, where 512 is the maximal number of input frames of the encoder considered in our implementation.  It has been found in \cite{hawthorne2021sequence} that such \texttt{WHICH-FRAME} tokens work better than the alternative \texttt{TIME-SHIFT} tokens \cite{oore2018time}.
   \item[SOS, EOS] [2 values] The start and end of a token sequence.
\end{description} 

The decoder and encoder are jointly optimized by the ``event loss,'' denoted as $L_\text{LM}$, calculated at the output of the decoder, formulated as the CE in the probability over all the possible tokens between the decoder output and the groundtruth at each time step of the decoder sequence.
The prior model \cite{hawthorne2021sequence} uses only $L_\text{LM}$ to train their whole model, while we use both: 
    $L_\text{total} = L_\text{AM} + L_\text{LM}$.

\begin{table*}
  \begin{center}
    \begin{tabular}{|lrrr|cc|cc|}
    \cline{5-8}
    \multicolumn{4}{c|}{} & \multicolumn{4}{c|}{\textbf{DI} ~~(Section \ref{sec:exp:di})} \\
    \hline
       \multirow{2}{*}{Model} &  \multirow{2}{*}{Batch size}  & \multirow{2}{*}{Sequence length}  & \multirow{2}{*}{Parameters}  
       & \multicolumn{2}{c|}{(Encoder output)}   & \multicolumn{2}{c|}{(Decoder output)}  \\
       & & & &  Onset F1 &  Frame F1 & Onset F1 &  Frame F1   \\
      \hline\hline
      Onset and Frame (OAF) \cite{hawthorne2017onsets} & 192  & 256 & 0.74M & 0.591 & 0.583   & ---  &  --- \\ 
      \hline
          & 32  & 256  & 4.58M & --- & ---  &  0.543 &  0.523\\
      CE-only Transformer \cite{hawthorne2021sequence}   
          & 128  & 128 & 4.58M  & --- &  --- &  0.554 &  0.524\\
          & 256  & 64  & 4.58M & --- & ---  &  0.568 &  0.537\\
      \hline
                & 32  & 256 & 6.01M  & 0.598 &  0.579  &  0.507 &  0.495\\
      Proposed multi-loss Transformer      
                & 128  & 128 & 6.01M  & 0.604 &  0.573 &  0.515 &  0.493\\
                & 256  & 64 & 6.01M  & \textbf{0.613} &  \textbf{0.582} &  0.514 &  0.496\\
      \hline
    \end{tabular}
    \caption{Transcription accuracy of various models trained on the training split of the DI recordings of EGDB (i.e., ``EGDB-DI'') and tested on the test split of EGDB-DI (i.e., same timbre). For the proposed model, we evaluate the result of both the encoder output and decoder output (since both are legitimate transcription results). We highlight the best result in bold.}
    \label{tab:di}
  \end{center}
\end{table*}

\begin{table}
  \begin{center}
    \begin{tabular}{|l|rr|rr|}
      \cline{2-5}
       \multicolumn{1}{c|}{} & \multicolumn{2}{c|}{\textbf{4,~5-th Amp}} & \multicolumn{2}{c|}{\textbf{Real-world Data}} \\
       \multicolumn{1}{c|}{} & \multicolumn{2}{c|}{(Section \ref{sec:exp:tone})} & \multicolumn{2}{c|}{(Section \ref{sec:exp:real})} \\
       \hline
      Model & Onset F1 &  Frame F1  & Onset F1 &  Frame F1 \\
      \hline\hline
      \cite{hawthorne2017onsets} : DI       &  0.568 &  0.536 & 0.524 & 0.514 \\
      \cite{hawthorne2017onsets} : 3Amps       &  0.588 &  \textbf{0.546} & 0.541 & 0.562 \\
      \hline
      \cite{hawthorne2021sequence} : DI    &  0.503 &  0.478 &  0.513 & 0.492\\
      \cite{hawthorne2021sequence} : 3Amps    &  0.532 &  0.523 & 0.521 & 0.484 \\
      \hline
      Ours : DI     &  0.534 &  0.512 & 0.527  & 0.561\\
      Ours : 3Amps    &  \textbf{0.592} &  0.526 &  \textbf{0.550} & \textbf{0.586}\\
    \hline
    \end{tabular}
    \caption{Transcription accuracy on testing data with unseen timbres (left: the test split rendered with the two held-out Amps; right: real-world recordings) by the three models trained on either the training split of EGDB-DI or EGDB-3Amps. We report the result of \cite{hawthorne2017onsets} and our model at the encoder side, and that of \cite{hawthorne2021sequence} at the decoder side.}
    
    \label{tab:3amp-and-real}
    \vspace{-3mm}
  \end{center}
\end{table}


\subsection{Implementation Details}

We split our dataset by 8:1:1 to create disjoint training, validation and test splits.
We implement the proposed model and the two baselines \cite{hawthorne2017onsets,hawthorne2021sequence} by our own in PyTorch.
The input to all the models are log-scaled mel-spectrograms computed by STFT with 2,048-pt Hann window, 512-pt hop size (each frame lasts for 11ms under 44.1 kHz audio), and 229 mel-filters.
We train all the models with the proposed dataset on an NVIDIA Tesla V100 GPU (with 32 GB memory) using the Adam optimizer \cite{kingma2014adam} with a learning rate (LR) of 1e--4,  trained until the onset validation score converges. 
All the models converge in less than half a day.
To the best of our knowledge, this represents the first attempt documenting the performance of the two baseline models for guitar transcription.

Unlike \cite{hawthorne2021sequence}, which used eight Transformer stacks for both the encoder and decoder,  we adopt only four stacks each for both our model and the CE-only baseline, because of the smaller size of our dataset and memory constraint of the GPU. We set the embedding size $d_\text{model}$ to 256, hidden layer size $d_\text{hidden}$ to 256, and use 8-head attention. Following  \cite{hawthorne2021sequence}, We use fixed absolute positional embedding for both the encoder and decoder. 

For convenience, we fix the length of the source sequence (i.e., encoder input; the number of input frames) and target sequence (i.e., decoder output; the number of tokens corresponding to that input audio segment) to be the same, zero-padding at the back when needed. 
According to our dataset, the token sequence corresponding to a 256-frame audio segment (roughly 3 seconds) has on average 61.61 tokens, or 17.00$\pm$10.15 notes. 
We explore the effect of sequence length and batch size  empirically in Section \ref{sec:exp:di}. 

\section{Experiment}
\label{sec:exp}

We use the \texttt{mir\_eval} library \cite{raffel2014mireval} to compute the ``onset F1'' and ``frame F1'' scores of each model. For onset F1, we consider an onset estimate as correct when it is within a $\pm$50ms tolerance window of a ground truth onset (following \cite{hawthorne2017onsets}). We do not evaluate the offsets.

\subsection{Evaluation on DI Recordings}
\label{sec:exp:di}
We firstly consider the case where both the training and testing splits are DI recordings (or, ``EGDB-DI''), therefore of the same timbre.
Table \ref{tab:di} shows that both OAF and the proposed multi-loss Transformer model outperform the CE-only Transformer model, possibly because our training set is not large enough to train a reliable language model for the decoder, necessitating the use of loss terms at the encoder output side.
We also see that, the Transformer models perform better using a larger batch size rather than a longer sequence length. The best result is obtained by the multi-loss Transformer model with batch size 256, using the output of its encoder.
Therefore, in what follows, we evaluate the accuracy of the encoder output rather than decoder output of our model.


\subsection{Evaluation on Unseen Timbres Rendered with Amps}
\label{sec:exp:tone}
Secondly, we evaluate the effect of timbre on the performance of guitar transcription. This time, we use the recordings of the test split rendered with the last two Amps (i.e., `iv' and `v' in Section \ref{sec:dataset:timbre}) for testing; and recordings of the training split of either the DI or those rendered with the first three Amps (or, ``EGDB-3Amps,'' which is three times larger than EGDB-DI) for training.
The following observations can be made from the left-hand side of Table \ref{tab:3amp-and-real}. First, for all the three models, training on ``3Amps'' works better than training on ``DI,'' demonstrating the advantage of having a multi-timbre dataset.
Second, if we use only EGDB-DI for training, we see that the onset F1 drops from 0.613 to 0.534 when we change the test set from DI to 4,~5-th Amp.
Third, using EGDB-3Amps for training, our model achieves the best onset F1 of 0.592, while \cite{hawthorne2017onsets} attains the best frame F1 score. 
We plan to do further error analysis in the future to gain insights into the behavior of the models.

\subsection{Evaluation on Real-world Recordings}
\label{sec:exp:real}
Similar observations hold in our final evaluation, where we test the models on five real-world recordings downloaded from YouTube and manually annotated by our musician.
The real-world recordings involve the use of pedals for effects such as delay and reverb, that are not seen in our data.
Moreover, as often the case, the real-world recordings are not DI signals and can involve the use of guitar types and Amps not seen in our data.
While the performance of all the models drops as expected, as shown in the right-hand side of Table \ref{tab:3amp-and-real} e.g., the best onset F1, as achieved by the proposed model, drops to 0.550), training on 3Amps is still advantageous. 
This adds empirical support of the importance of using a timbre-rendered dataset.



\begin{figure}[t]
    \centering
    \includegraphics[width=.48\textwidth]{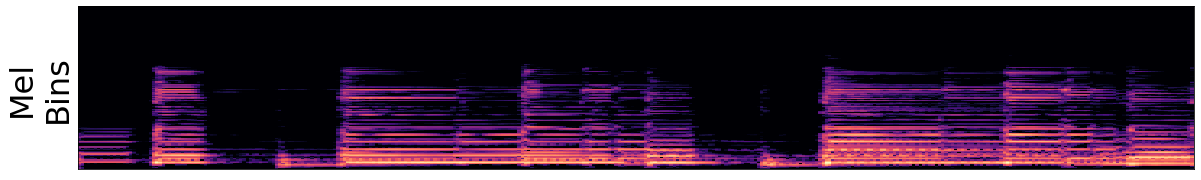} \\
    \includegraphics[width=.48\textwidth]{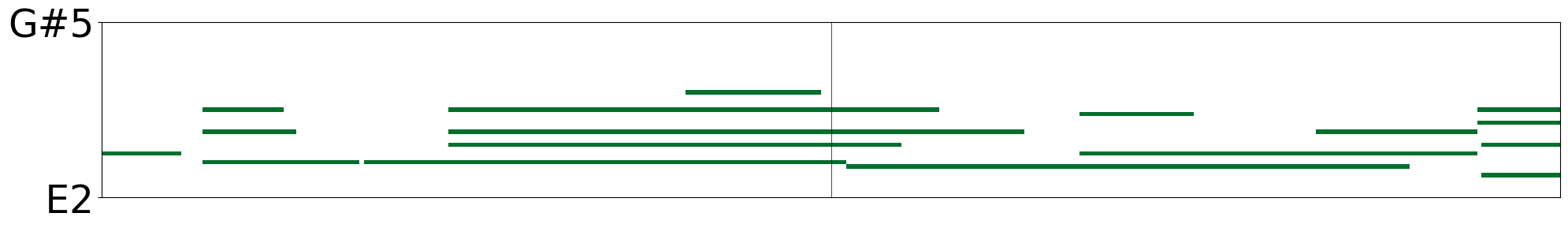}
    \includegraphics[width=.48\textwidth]{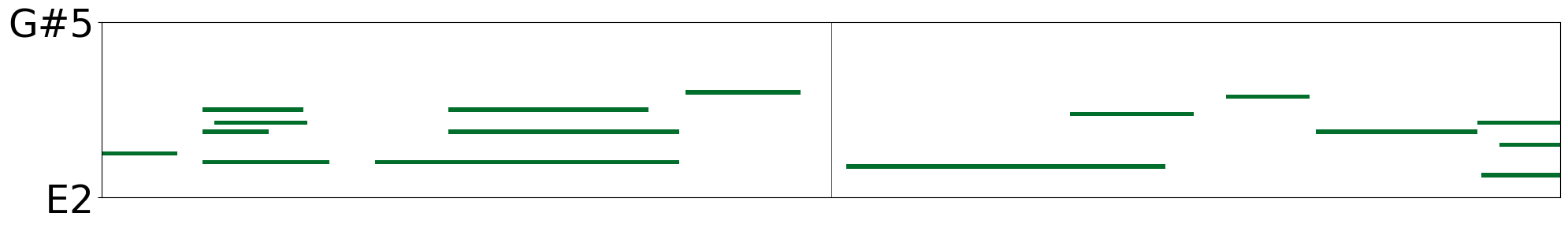}
    \caption{(Top) the spectrogram of a real-world testing recording, (middle) human transcription, (bottom) transcription result of the proposed multi-loss Transformer model, trained on EGDB-3Amps.}
    \label{fig:generatd_result_to_tab}
\end{figure}

The real-world recordings and the transcription result of our  model can be found at the demo website.
The result of one such recording is illustrated in Figure \ref{fig:generatd_result_to_tab}.

\section{Conclusion}
\label{sec:conclusion}

In this paper, we have presented a new data collection methodology and a resulting new public dataset of multi-timbre electric guitar. 
We have also presented a multi-loss seq2seq Transformer model for AMT, and benchmarked its performance along with another two existing models \cite{hawthorne2017onsets,hawthorne2021sequence} using our dataset and a small collection of real-world recordings. 
The result of the best model in onset F1 falls within 0.550--0.613 across different test sets, while the onset F1 of state-of-the-art models for piano transcription (e.g., \cite{hawthorne2021sequence}) tends to be higher than 0.950 on MAESTRO \cite{hawthorne2018enabling}. 
This prompts future work to close the performance gap between piano  and guitar transcription.



\vfill
\pagebreak


\bibliographystyle{IEEEbib}
\bibliography{strings,refs}

\end{document}